\documentclass[12pt,preprint]{aastex}

\usepackage{psfig,natbib}
\usepackage{palatcm}

\def\fel{\hbox{f$_{\rm el}$}}

\def\la{\mathrel{\hbox{\rlap{\hbox{\lower4pt\hbox{$\sim$}}}\hbox{$<$}}}}
\def\ga{\mathrel{\hbox{\rlap{\hbox{\lower4pt\hbox{$\sim$}}}\hbox{$>$}}}}

\newcommand{\be}{\begin{eqnarray}}
\newcommand{\ee}{\end{eqnarray}}

\newcommand{\msol}{\ifmmode{{\rm M}_\odot}\else{M$_\odot$}\fi}
\newcommand{\foe}{\ifmmode{10^{51}}\else{$10^{51}$}\fi}

\newcommand{\xni}{\ifmmode{{\rm X}_{\rm Ni}}\else{X$_{\rm Ni}$}\fi}
\def\ang{\hbox{\AA}}
\def\Teff{\ifmmode{T_{\rm eff}}\else{\hbox{$T_{\rm eff}$} }\fi}
\def\Rzero{\ifmmode{R_0}\else{\hbox{$R_0$} }\fi}

\def\SP2{{\tt IBM SP2}}

\def\PC2{{\tt PC$^2$}}

\def\logg{\log(g)}
\def\mh{[{\rm M/H}]}

\def\inu{\ifmmode{I_{\nu}}\else{\hbox{$I_{\nu}$} }\fi}
\def\snu{\ifmmode{S_{\nu}}\else{\hbox{$S_{\nu}$} }\fi}
\def\jnu{\ifmmode{J_{\nu}}\else{\hbox{$J_{\nu}$} }\fi}

\def\fep{\ifmmode{{\rm Fe II}}\else\hbox{Fe~II }\fi}

  at 12.0 true pt 

\def\phoenix{{\tt PHOENIX}}

\def\water{{H$_2$O}}

\def\phoenix{{\tt PHOENIX}}

\def\water{{H$_2$O}}
\def\jtio{SCAN-TiO}
\def\amestio{AMES-TiO}

\def\mtwater{MT-\water}
\def\ameswater{AMES-\water}

\def\b{\beta}
\def\m{\mu}

\def\rout{\ifmmode{r_{\rm out}}\else\hbox{$r_{\rm out}$}\fi}
\def\tmax{\ifmmode{\tau_{\rm max}}\else\hbox{$\tau_{\rm max}$}\fi}
\def\tstd{\ifmmode{\tau_{\rm std}}\else\hbox{$\tau_{\rm std}$}\fi}
\def\vmax{\ifmmode{v_{\rm max}}\else\hbox{$v_{\rm max}$}\fi}
\def\muE{\ifmmode{\mu_{\rm E}}\else\hbox{$\mu_{\rm E}$}\fi} 
\def\pE{\ifmmode{p_{\rm E}}\else\hbox{$p_{\rm E}$}\fi} 
\def\bmax{\ifmmode{\b_{\rm max}}\else\hbox{$\b_{\rm max}$}\fi}

\def\ang{\hbox{\AA}}

\def\Msun{\hbox{$\,$M$_\odot$} }

\def\Teff{\hbox{$\,T_{\rm eff}$} }

\def\rout{\hbox{$r_{\rm out}$} }

\def\chistd{\ifmmode{\chi_{\rm std}}\else\hbox{$\chi_{\rm std}$}\fi}

\def\K{\,{\rm K}}
\def\msol{$M_\odot$}
\def\foe{10^{51}}

\def\xni{{\rm X}_{\rm Ni}}

\def\teff{{\rm T}_{\rm eff}}

\def\lstar{\ifmmode{\Lambda^*}\else\hbox{$\Lambda^*$}\fi} 
\def\Rop{\ifmmode{[R_{ij}]}\else\hbox{$[R_{ij}]$}\fi}

\def\Rji{\ifmmode{[R_{ji}]}\else\hbox{$[R_{ji}]$}\fi}
\def\Rstar{\ifmmode{[R_{ij}^*]}\else\hbox{$[R_{ij}^*]$}\fi}

\def\Rjistar{\ifmmode{[R_{ji}^*]}\else\hbox{$[R_{ji}^*]$}\fi}
\def\DRji{\ifmmode{[\Delta R_{ji}]}\else\hbox{$[\Delta R_{ji}]$}\fi}
\def\DRij{\ifmmode{[\Delta R_{ij}]}\else\hbox{$[\Delta R_{ij}]$}\fi}

\def\ns{\ifmmode{N_{\rm s}}          
        \else\hbox{$N_{\rm s}$}\fi}

\def\mat#1{{\bf #1}}     
\def\vek#1{{#1}}         

\newcount\eqcount
\def
  \nummer{
    \global\advance\eqcount by 1
    (\the\eqcount)
  }

\def
  \numadv{
    \global\advance\eqcount by 1
  }

\def
   \numout#1{
     (\the\eqcount #1)
  }

\def\ivek#1#2{\ifmmode{\vek{I}^{#1}_{#2}}
        \else\hbox{$\vek{I}^{#1}_{#2}$}\fi}

\def\tmat#1#2{\ifmmode{\mat{t}^{#1}_{#2}}
        \else\hbox{$\mat{t}^{#1}_{#2}$}\fi}
\def\rmat#1#2{\ifmmode{\mat{r}^{#1}_{#2}}
        \else\hbox{$\mat{r}^{#1}_{#2}$}\fi}
\def\bvek#1#2{\ifmmode{\beta^{#1}_{#2}}
        \else\hbox{$\beta^{#1}_{#2}$}\fi}

\def\lp{\ifmmode{\lambda^+_\tau}           
        \else\hbox{$\lambda^+_\tau$}\fi}
\def\lm{\ifmmode\lambda^-_\tau             
        \else\hbox{$\lambda^-_\tau$}\fi}

\chardef\tilt=126
\begin{document}
\bibliographystyle{apj}

\title{TiO and H$_2$O absorption lines in cool stellar atmospheres}

\author{France Allard}
\affil{C.R.A.L (UML 5574) Ecole Normale Superieure, 69364 Lyon Cedex 7,
rance\\
E-Mail: \tt fallard@ens-lyon.fr}
\author{Peter H. Hauschildt}
\affil{Dept.\ of Physics and Astronomy \& Center for Simulational Physics,
University of Georgia, Athens, GA 30602-2451\\
Email: {\tt yeti@hal.physast.uga.edu}}
\and
\author{David Schwenke}
\affil{NASA Ames Research Center, Mail Stop 230-3, Moffett Field,
CA 94035-1000\\
E-mail: {\tt schwenke@pegasus.arc.nasa.gov}
}

{\em ApJ, in press. Also available at\\
\tt ftp://calvin.physast.uga.edu/pub/preprints}

\begin{abstract}

We compare the structures of model atmospheres and synthetic spectra calculated
using different line lists for TiO and water vapor. We discuss the effects of
different line list combinations on the model strutures and spectra for both
dwarf and giant stars.  It is shown that recent improvements result in
significantly improved spectra, in particular in the optical where TiO bands
are important. The water vapor dominated near-IR region remains problematic as
the current water line lists do not yet completely reproduce the shapes of the
observed spectra.  We find that the AMES TiO list provides more opacity in most
bands and that the new, smaller oscillator strengths lead to  systematically
cooler temperatures for early type M dwarfs than previous models.  These
effects combine and will help to siginificantly improve the fits of models to
observations in the optical as well as result in improved synthetic photometry of M
stars. We show that the \cite{davis86} \fel-values for the $\delta$ and
$\varphi$ bands of TiO best reproduce the observed (V-I) color indices. 


\end{abstract}

\section{Introduction}

Over the past decade, model atmospheres and synthetic spectra for late-type
stars have improved hand-in-hand with higher quality opacities.  In 1994,
quality lists of transitions of the water vapor molecule based on ab initio
molecular calculations became available \cite[]{UCL,schryb94,H2OJorg} which
allowed the computation of the first Direct Opacity Sampling (hereafter dOS)
model atmospheres for late-type dwarfs \cite[]{h2olet} and brown dwarfs
\cite[]{gl229b}, later to become the NextGen models described in
\cite[]{ng-hot}.  Showing a more physical description of their main opacities,
the NextGen dOS model atmospheres promised a better description of the Spectral
Energy Distribution (hereafter SED) of cool stars.  And this appeared to be
verified for the infrared SED of M dwarfs \cite[]{Jones96,leg96,araa}.

But despite these fundamental improvements, the NextGen models have
failed to match adequately several of the optical (spectroscopic and
photometric) properties of late type dwarfs and giants. In fact, the
dOS models \cite[]{B95,h2olet,araa,ng-hot} provided a worse fit to the
optical SED of lower main sequence dwarfs than previous models based
on simplified molecular opacities \cite[][hereafter, AH95]{MDpap}.
The models could not reproduce the tight relation formed by M dwarfs
in the V-R versus R-I two-color diagram, indicating a systematically
shallower slope of the optical SED (defined by TiO absorption) then
observed in these stars.  A systematic flux excess in the spectral
region sampled by the V bandpass ($0.4-0.65 \m$m) was noted as well in
dwarfs as in red giants.  \cite{BCAH98} observed that this excess in
the NextGen dwarf models translated into lower main sequence
isochrones deviating progressively to the blue (by up to 1.0
magnitude!) in (M$_V$,V-I) color-magnitude diagrams, for masses lower
than about 0.5 M$_\odot$ (T$_{\rm eff}\leq 3800\,$K).  \cite{BCAH97}
examined a variety of globular clusters and showed that this departure
of the models decreased in amplitude with decreasing metallicity. The
problem seemed therefore confirmed to be caused by a lack of opacity
of an oxygen compound\footnote{Hydride absorption bands only get
stronger relative to the continuum with decreasing metallicity over
the range covered by the globular clusters studied in \cite{BCAH97} in
the optical spectral range.}.

Only three independent models of the TiO molecule and corresponding
lists of transitions were available so far to the construction of
model atmospheres.  The first model was constructed over two decades
ago by \cite{coll75} and was restricted by the computational
limitations of the time.  The Collins line list was intended to model
the extended atmospheres of red giants and did not include high energy
and otherwise weak transitions important by their number in the hotter
environments of red dwarf atmospheres.  It also neglected the red
$\epsilon$ system of TiO.  \cite{TiOJorg} extended Collins' work to
TiO isotopic transitions, included the $\epsilon$ system from revised
molecular rotational constants, and adopted the laboratory oscillator
strengths of \cite{davis86}.  It is therefore understandable that the
resulting limited list of transitions caused shortcomings in the
NextGen model atmospheres.  The second TiO list was constructed by
\cite{cdrom15} and is used in his ATLAS9-12 atmospheres.  The third
model was constructed by \cite{Plez92} using also the \cite{davis86}
oscillator strengths, and is used in his version of the MARCS
atmosphere code.  All three independent models yielded the visual flux
excess in different proportions.  \cite{Plez98} suggested that the
missing opacity is due to missing TiO band systems in current lists,
and added the TiO a-f system at 0.5 $\m$m to his list. However, this
helped him only partially to resolve the V-band flux excess problem.

Recently, \cite{Langhoff97} constructed a new model of the TiO
molecule and published new lifetimes and oscillator strengths that
improved significantly upon the 1986 values of Davis et al.\
\cite[]{Valenti97,Plez98}.  \cite{ames-tio} has subsequently computed
a corresponding list of transitions complete to the high energies and
therefore more suitable for general model atmosphere applications.  In
this paper we present the results of including this new TiO line list
as well as the new \ameswater\ list
in the construction of model atmospheres and synthetic spectra for
late-type dwarfs and red giants.

\section{Model calculations}

We have calculated the models presented in this paper using version 10.3 of our
general model atmosphere code \phoenix. Details of the code and the general
input physics are discussed in \cite{ng-hot} and references cited therein.  The
models for M giants were calculated with the same setup, however, they employ
spherical geometry (including spherically symmetric radiative transfer).  For
giant models with low gravities ($\logg \le 3.5$), this can be an important
effect for the correct calculation of the structure of the model atmosphere and
the synthetic spectrum \cite[]{epscma,betacma}.  The main difference between
the models presented in \cite{ng-hot} and the models presented here is the use
of the new AMES line lists for \water\ \cite []{ames-water-new} and
TiO \cite[]{ames-tio}, but we have also adjusted the empirical oscillator
strengths of VO and CaH absorption bands to respect their strength relative to
TiO bands (note that VO and CaH absorption is still treated in the Just
Overlapping Line Approximation due to lack of adequate line data).  Our
combined molecular line list includes about 500 million molecular lines.  These
lines are treated with a dOS technique where each line has its individual Voigt
(for strong lines) or Gauss (weak lines) line profile (in the standard OS method
tables of precomputed opacities are used).  They are selected for
every model from the master line list to the beginning of each model iteration
to account for changes in the model structure, see \cite{ng-hot} for details.
This procedure selects about 215 million molecular lines for a typical giant
model with $\Teff\approx 3000\K$ and about 130 million molecular lines for a
dwarf model with the same effective temperature.  Therefore, we generally use
the parallelized version of phoenix\ \cite[]{parapap,parapap2,jcam} to perform
the calculation efficiently on parallel supercomputers.  Details of the TiO and
H$_2$O lists are given in the next subsections.

\subsection{Water lines}

The effects of water lines on the M dwarfs SED was discussed in
\cite{h2olet}.  For the work presented here, we have replaced the UCL
water vapor line-list \cite[][hereafter: \mtwater]{UCL,schryb94} used
in \cite{ng-hot} with the AMES water line-list \cite[][hereafter:
\ameswater]{ames-water-new}.  This list includes about 307 million
lines of water vapor.  For the calculations shown in this paper we
have used H$_2 ^{16}$O and neglected other, much less abundant, isotopes
of this molecule.

The water vapor opacity is governed by the completeness of the line
list used, but also by the adopted atomization energy.  The partition
function of the molecule cancels out in the final absorption
coefficient, after we have multiplied cross-sections by number
densities.  But since water is an important chemical equilibrium
specie, errors in the partition function can affect indirectly the
model structure and spectra.  The AH95 models were based on the
\cite{ludwig} hot flames water cross-sections in the form of
straight means, and used the JANAF partition function for water vapor
\cite[]{irwin88}.  The NextGen models where, on the other hand,
computed with the \mtwater\ line list and a partition function computed
from the \mtwater\ levels.  We note that the \ameswater\ partition
function is practically identical to JANAF values, while the \mtwater\
value is smaller than JANAF for temperatures above 3000K, possibly due to the
energy levels missing in the \mtwater\ data.  We have therefore adopted for this and later
work the JANAF partition function.  We use an atomization energy of
$9.5119\,$eV from \cite{irwin88} for all models since AH95.

\subsection{TiO lines}

The main point of this paper is the comparison of the model structure and the
synthetic spectra obtained by using the list of TiO lines from
\cite[][hereafter: \jtio]{TiOJorg} and the new list of TiO lines from
\cite[][hereafter, \amestio]{ames-tio}.  The \amestio\ list includes a total of
about 172 million lines, about 44.6 million of these are for the most abundant
isotope $~^{48}$TiO and about 32 million lines for each of the remaining 4
isotopes $^{46,47,49,50}$TiO).  But beyond the completeness of the line-list,
two more considerations affects the overall opacity produced by TiO, and
explain systematic differences between model versions and by different authors:
the atomization energy ($D_0^0$) determines the number density of TiO, and the
TiO band oscillator strengths\footnote{$\fel= f_{\nu'\nu''}/q_{\nu'\nu''}$,
where the $f$'s are the oscillator strengths and the $q$'s are the
Franck-Condon factors of the transition $\nu'\nu''$} have been derived from
sunspot observations \cite[hereafter: DLP86]{davis86}, laboratory experiments
\cite[hereafter: HNC95]{TiOfel95}, as well as from ab initio calculations
\cite[hereafter L97]{Langhoff97}.  \cite[ hereafter:B90] {brett90} derived
astrophysical \fel\ values by fitting the optical SEDs of red giants, using an
atomization energy of $7.76\,$eV.  He quoted that reducing this value by
$0.3\,$eV would increase his \fel\ by a factor 2.5.  The most recent estimate of
D$_0^0$ for TiO is now $6.92\,$eV, which suggests that the B90 \fel\ values are
underestimated by as much as a factor 7!  We summarize in table \ref{feltab}
the various sources of oscillator strengths available for TiO.

The models of \cite{allardphd} used \fel\ values from B90 together
with the straight-mean TiO opacities by \cite{coll75} and
\cite{tio74}, and assuming an atomization energy of $6.87\,$eV.
The first comparison of these models to
the SED of M dwarfs \cite[]{kirk93} revealed the inadequacy of this
combination of parameters for TiO which produced far too weak optical
opacities.  We have therefore, since the AH95 model series, employed
the updated value of $6.92\,$eV,
together with the larger laboratory \fel-values of DLP86.  These two
modifications combined to significantly increased the strength of TiO
opacities in the models, bringing the AH95 and later the dOS NextGen
models in improved agreement with the SED of M dwarfs.  Any differences
in the predictions of the AH95 and NextGen models are therefore purely
due to the opacity technique (Straight-Mean versus dOS) and to the
completeness of the line-list used.  The incompleteness of the \jtio\
line-list allows photons to escape between absorption bands \cite[see
e.g.][]{Valenti97}, and thus leads to systematically and increasingly
(with higher $\Teff$) bluer optical colors (V-I) than observed
\cite[]{solar-evol,clusterpap}.  For the current models we therefore
explore the use of the more complete \amestio\ line-list, and the yet
larger theoretical \fel-values of L97.

\section{Results}

We have calculated a number of model atmospheres using either the
\jtio\ or the \amestio\ list of TiO lines and using either \ameswater\
or \mtwater\ as source of the \water\ lines.  All the other input
physics is the same for both sets of models.  All models have been
fully converged with their respective set of parameters.  Note that
these models have been constructed for the purpose of this paper only
and not to model individual stars and thus do not include dust
formation and opacities which is important in atmosphere models with
effective temperatures below about $2500\K$, such models are presented
in a subsequent paper.  In the following, we will discuss the results
for the dwarf and giant models separately.  The baseline for our
comparisons are the NextGen models \cite[]{ng-hot} for the dwarfs and
the NG-giant models \cite[]{ng-giants} for the giants.

\subsection{Effects of different TiO line lists}

The models discussed in this section were all calculated using
\ameswater\ to isolate the effects of different TiO line lists on the
model spectra and structures.

\subsubsection{M dwarf models}

In Figs.~\ref{dwarfs-blue} and \ref{dwarfs-red} we show a comparison
of model spectra calculated with \amestio\ (full curves) and with
\jtio\ (dotted curves) (both using our adopted \fel-set, as quoted in table
\ref{feltab}) for several effective temperatures.  The gravity
($\logg=5.0$) and abundances (solar) were selected to be
representative of M dwarfs in the solar neighborhood.  In both
figures, the resolution of the synthetic spectra was reduced by boxcar
smoothing to $20\ang$.  At high effective temperatures, the two sets
of models are nearly identical due to reduced importance of TiO
absorption.  At very low $\Teff$ the two line lists apparently agree
very well since only the lowest levels of TiO remain populated.  It is
essentially between $\Teff \approx 2000\K$ and $\approx 3500\K$ that
the largest completeness and quality effects of the TiO line lists are
seen.

Fig.~\ref{tio29} indicates the location of each TiO band system for a
2900K model.  From this it becomes clear that the addition of a-f
transitions, which depress the continuum from 0.4 to 0.5$\m$m, is one
of the largest improvements brought by the \amestio\ list to our
models. 
We note that the entire optical regime from 0.4 to 0.75$\m$m
shows generally {\bf more} opacity in the \amestio\ models then using
the \cite{TiOJorg} (hereafter: J94) line list.  The $\epsilon$ bands at 0.82 to 0.88$\m$m have a
more precise shape in the \amestio\ list, and come out stronger as
well.  This is a result of the completeness of the \amestio\ list
which also removes flux excess escaping between the troughs of the 
bands.  We note however that
some regions, such as in the $\gamma$ band near 0.78$\m$m, show less
opacity in the new models.  

The main effect of the new \amestio\ on spectroscopic and
photometric $\Teff$ estimates will however be dominated by the change
we make to the oscillator strengths.  The L97 \fel\ values being
generally smaller than the DLP86 values adopted by J94, models of
early-type M dwarfs using the new \amestio\ setup should predict
systematically lower effective temperatures then did prior models
(NextGen, AH95, etc., see also Fig.\ref{TiO3500} below).  And beyond
the enhanced completeness of the \amestio\ list to high temperature
transitions, the need for a cooler model should also contribute to
making the TiO bands fit better a given star i.e. larger bands with
less flux escaping from deeper, hotter atmospheric regions between
them.

We could have opted to use the HNC95 laboratory \fel-values as did
\cite{Plez98}, but since the L97 ab initio values agree quite well, 
we decided to keep these, except for the $\delta$ and $\varphi$ band
systems. 
The reason the oscillator strengths for the $\delta$ and $\varphi$ bands
are less accurate is that it is very hard to get a good description of
the b state, which is the upper state in both bands.
For the $\delta$ system, L97 derives an oscillator strength
which is, as opposed to all other bands, twice as large as the DLP86
value.  And the $\delta$ and $\varphi$ \fel-values cannot be
corroborated by recent experimental values.  Such a strong $\delta$
band system would be difficult to bring in agreement with M dwarfs
observations.  Indeed, prior models have all shown a gradually
increasing departure to the blue of the main sequence in M$_V$ vs
$V-I$ diagrams \cite[]{baraffe95,solar-evol}.  Such departure is
significantly improved using the new TiO list if one keeps a weak
$\delta$ band as indicate preliminary results of evolution models to
be published separately (see also Fig.~\ref{VRI} below).  We have therefore adopted to keep the DLP86
oscillator strength values for the reddest two TiO bands until new
laboratory experiments can either confirm or infirm the L97
predictions.  The summary of our adopted set of oscillator strengths
for TiO is presented in Table~\ref{feltab}.

\subsubsection{M giant models}

The results for the giant models are similar to the results for the
dwarfs.  Figures \ref{giants-blue} and \ref{giants-red} show synthetic
spectra for 3 representative giant models with the indicated effective
temperatures.  The models have in common the parameters $\logg=0.5$,
$M=5\Msun$ and solar abundances.  The differences between the
\amestio\ (full curves) and \jtio\ (dotted curves) models are somewhat
larger for giants than for the dwarfs in the blue spectral region due
to an increased sensitivity to the added a-f system opacities in the
\amestio\ line list.  It is however somewhat less pronounced in the
red spectral region where TiO bands are weaker in giants.  The
``spikes'' that are apparent in the \jtio\ spectrum with
$\Teff=3000\K$ are absent in the \amestio\ models.  These spikes were
one of our major problems in fitting observed spectra of giants.  For
larger $\Teff$ the differences between the spectra diminishes quickly
as TiO becomes less important in the giants.  This happens at lower
effective temperatures compared to the dwarfs because of the lower
pressures in giant atmosphere which results in smaller partial
pressure of molecules as compared to dwarfs.

\subsubsection{Model Structures}

A comparison of the model structures for both dwarf and giant model
reveals only very small differences between structures calculated with
\amestio\ and \jtio.  We plot the differences in electron temperature as
well as the relative differences between the \amestio\ and \jtio\
models for dwarfs and giants in Figs.~\ref{dwarf-structure} and
\ref{giant-structure}, respectively.  The changes are generally very
small, only in the per-cent range for the gas pressures and about
$10\K$ maximum difference between the electron temperatures for the
dwarf models. For the giant models the differences are somewhat larger.
The changes in the opacity averages are generally small but largest
for the Rosseland mean opacity in the outer layers of the giant
models.  The temperatures are higher in \amestio\ model for both the
giant and the dwarf models, however, the gas pressures are lower in the
\amestio\ dwarf model but higher in the \amestio\ giant model.  Overall
the changes are modestly small, indicating that the detailed effects of
the TiO line lists do not have a large effect on the model structure
itself.

\subsection{Effects of different water vapor line lists}

In figure \ref{mt-compare} we show the effects of different water line
lists on the synthetic spectra for M dwarfs.
All models shown in the graph otherwise use the same
line-lists (\amestio\ was used for the TiO lines).  In overall, we can
see that the changes in the water vapor line lists are of larger
amplitude than the changes in the completeness of the TiO line list.
The models calculated with \ameswater\ show a totally different shape
of the $1.4\,\mu$m band, both weaker and wider then predicted by the
\mtwater\ model.  The completeness of the new line list to high
temperatures helps block more flux escaping from deeper, hotter layers
of the models around $1.6\,\mu$m, and $2.2\,\mu$m.  This promises a much
better description of observations in general.  

We also find important changes of the model structure as shown in
Fig.\ref{mt-structures}.  The differences of the electron temperatures
can reach $100\K$, the gas pressures can differ by $20\,$\% and the
opacity averages, in particular the Rosseland mean, can differ by
close to $60\,$\%.  These effects are much larger in the outskirts of
the atmosphere than the changes in the structures caused by different
TiO line lists (see Fig.~\ref{dwarf-structure}).  As a result, the use
of the \ameswater\ line list also affects the optical spectra, causing
weaker TiO bands than obtained with the \mtwater\ line list.  Models
of early-type M dwarfs based on the \ameswater\ line list therefore
systematically predict yet lower effective temperatures for a given
star.

\subsection{Combined effects}

In Figs.~\ref{TiO3500} and \ref{H2O3500} we display a comparison
between NextGen models (which use \mtwater\ and \jtio) and models that
use the \ameswater\ and \amestio.  The wavelength range of important
filters and band identifications for TiO are given on the figures.
The TiO bands in the ``AMES-atmosphere'' are considerably weaker than
those of the NextGen model spectrum as a result of the smaller
oscillator strengths used, and the structural effects.  On the other
hand, the water bands are stronger in the AMES-atmosphere than in the
NextGen model.  This model has a relatively high temperature, thus the
higher energy levels of the water molecule are relatively more
important than for models with lower $\teff$ (however, the
concentration of water molecules is reduced so the overall water
opacity is smaller than in the cooler models).  

To better judge of the impact of these opacity changes on the overall SED of M
dwarfs in general, we have computed synthetic photometry as described in AH95
for three sets of models: (1) the NextGen grid based on J94 and \mtwater\, (2)
the AMES grid based on \amestio\ and \ameswater\ opacities, and (3) the AMES-MT
grid based on \amestio\ and \mtwater\ opacities.  The results are compared to a
photometric sample of M dwarfs \cite[]{leggett92} in Figs.~\ref{VRI},
\ref{JHK}, and \ref{IJK}.  Since M dwarfs form a tight sequence in optical VRI
two-colors diagram despite the age and metallicity scatter of the sample (see
AH95), this diagram imposes a strong constraint on model atmospheres.  We find
that models based on \amestio\ opacities are systematically redder in $V-R$ and
$V-I$ than models based on the J94 line list.  The new models agree much better
with observations and the new TiO data removes most of the discrepancy shown by
the NextGen models in the lower main-sequence.  Small remaining discrepancies
may be attributed to the JOLA handling of VO and CaH which tend to overestimate
slightly their opacities in the present models.  \cite{lhs1070} already studied
the low resolution HST/FOS spectra of an M6 dwarf (LHS1070A) and found the
AMES-MT models indeed agree quite well both with the observed SED and absolute
fluxes within errors on the parallax of the system.  They however noticed that
some ``continuum'' flux excess remains important in the visual part of the SED
(0.45 to $0.65\,\mu$m).  However, there is no {\em a priori} reason to assume
that the a-f oscillator strengths are inaccurate, and these remaining problems
could be related to other effects on the model structure.

The use of \ameswater\ seems also to bring some improvements to the modeling of
near-infrared colors.  Fig.~\ref{JHK} shows that the late-type dwarfs can be
better reproduced by the new water opacities than by the \mtwater\ line list.
However this diagram is sensitive to both gravity (lower gravity models loop
lower) and metallicity, which makes it difficult to constrain the models on the
adequacy of the water opacities used with them.  \cite{leg98} and \cite{leg99} have
already used the AMES models in their analysis of M dwarfs and brown dwarfs,
and found an excellent general agreement of the predicted near-infrared SED
with observations.  However these analyses used the models to derive the
parameters of the studied stars and brown dwarfs based on fits to the
near-infrared SED or photometry, and could not make an independent  statement
on the quality of the water line list. 

M dwarfs form again a sequence in the mixed-colors IJK diagram
(Fig.~\ref{IJK}), although less tightly than in the VRI diagram.
Unresolved binary stars produce K-band flux excess and lie below the
sequence.  H$_2$ pressure-induced opacities depress the K-band flux of
metal-depleted dwarfs so that they systematically lie above the
sequence.  But, as opposed to the JHK diagram, this one is not
particularly sensitive to gravity in M dwarfs which allows a sequence
to be defined.  Models should pass therefore through the bulk of
early-type M dwarfs at $J-K=0.8$, and follow a relatively $J-K$-insensitive
sequence towards late-type dwarfs.  We find that models based on the
\ameswater\ opacities lie, as did the AH95 models before them, 0.2
magnitude in $J-K$ to the blue of the observed sequence!  And our
tests show that this result is independent of the TiO opacities used.
The NextGen models already reproduced perfectly the location of lower
main-sequence stars in this diagram.  And AMES-MT models computed
using the \amestio\ and \mtwater\ line lists behave adequately both in
the optical and infrared.  Why?  Perhaps the new water vapor line list
is still not complete enough to high temperatures and lacks opacity in
the $J$-bandpass i.e. around $1.3\,\mu$m?  Or would it have too much
opacity in the $K$-window i.e. around $2.2\,\mu$m?  Until these
questions can be answered, we hope that the two main grids of models
we have computed (AMES and AMES-MT) will allow independent detailed
confrontations to observations of cool stars that will locate more
precisely the source of the problem (e.g. Leinert et al., in
preparation).

\section{Summary and Conclusions}

A long standing problem with M dwarf models was that prior TiO line lists were
incomplete to high temperatures.  The use of ``straight means'' (AH95 models)
helped by the coarseness of the treatment to block flux which otherwise
escapes between lines in the incomplete list.  But these models also blocked
too much flux in most cases, and were only appropriate in late-type M dwarfs
when TiO bands are already very strong.  Clearly, a more complete line list was
needed to model stars from the onset of TiO formation to its gradual
disappearance from the gas phase in brown dwarfs.  The \amestio\ list now
serves beautifully this purpose.  We find that the list provides more opacity
in most bands and suppress adequately flux between bands.  The new, smaller
oscillator strength values also play an important role in systematically
assigning cooler models (at least for early type M dwarfs) to a given star,
this way contributing to broader bands and lesser inter-band flux as well.
These effects combine and should resolve most of the previously observed
discrepancy between models and observations in the optical SED and photometry
of M stars. \cite{lhs1070} note however that flux excess remains substantial in
the visual spectrum, suggesting some further incompleteness or \fel\
inaccuracies of the new TiO in the a-f system.

In order to better reproduce the observed (V-I) color indices, we had
to retain in the present models the \cite{davis86} \fel-values for the
two reddest band systems: $\delta$ and $\varphi$.  For these two band
systems, the theoretical estimates of \cite{Langhoff97} predicts an
unexpectedly large \fel-value ratio, while no laboratory estimate
\cite[]{TiOfel95} are available to corroborate this.  And we find, as did
\cite{AP98} in red giants for the $\delta$ band, that models based
upon the DLP86 \fel-values for these two bands reproduce adequately
their observed depths in M dwarfs.  

The introduction of the \ameswater\ opacities bring solid improvements of the
near-infrared SED of late-type dwarfs, but fails as the AH95 models did to
reproduce adequately the $J-K$ colors of hotter stars.  Water vapor is a more
important factor for the structure of the atmosphere than TiO because its
overall opacity is larger and its lines are closer to the peak of the SED than
the TiO bands, so the flux blocking effect of water vapor is more important for
the temperature structure than that of TiO opacities for these low
temperatures.  Schwenke and collaborators at NASA AMES are preparing a new
dipole moment function for H$_2$O, which may change the high temperature high
overtone water bands, and help resolve this discrepancy in the near future.  

Until a revised version of the \ameswater\ line list becomes available,
we have therefore generated two sets of model atmospheres for cool
stars which allow to investigate these issues: the AMES grid based on
the new TiO and H2O opacities, and the AMES-MT grid which rely on the
\amestio\ and \mtwater\ opacities. 


\acknowledgments 
We thank David Alexander for helpful discussions and the referee, U.G.
J{\o}rgensen, for his very helpful comments.  This work was supported in part
by grants from CNRS and INSU, NSF grant AST-9720704, NASA ATP grant NAG 5-3018
and LTSA grant NAG 5-3619 to the University of Georgia, and NASA LTSA grant
NAG5-3435 to Wichita State University.  This work was supported in part by the
P\^ole Scientifique de Mod\'elisation Num\'erique at ENS-Lyon.  Some of the
calculations presented in this paper were performed on the IBM SP2 of CNUSC,
the SGI Origin 2000 of the UGA UCNS, and on the IBM SP2 of the San Diego
Supercomputer Center (SDSC) with support from the National Science Foundation,
and on the Cray T3E of the NERSC with support from the DoE.  We thank all these
institutions for a generous allocation of computer time.

\clearpage

\bibliography{yeti,opacity,mdwarf,radtran,general,opacity-fa,local}

\clearpage
\section{Tables}

\begin{table}[ht]
\caption[]{\label{feltab}TiO \fel\ values ($\fel = f_{v1v2} / q_{v1v2}$)}
\begin{center}
\begin{tabular}{*{8}{r}}
\hline
\hline
 system   & lam0  &  B90  &  J94* & DLP86   & L97   & AP98 &adopted \\
\hline
  alpha  &$5170.7$ &  $0.10$  & $0.17$  & $0.106$ & $0.105$  &$ 0.106$
& $0.105$ \\
   beta  &$5605.2$ &  $0.15$  & $0.28$  & $0.125$ & $0.176$  &$ 0.125$
& $0.176$ \\
  gamma' &$6192.5$ &  $0.08$  & $0.14$  &$0.0935$ & $0.108$  &$0.0935$
& $0.108$ \\
  gamma  &$7095.8$ &  $0.09$  & $0.15$  &$0.0786$ & $0.092$  &$0.0786$
& $0.092$ \\
epsilon  &$8407.6$ &$0.0024$  &$0.014$  &$<0.006$ & $0.002$  &$0.0023$
& $0.002$ \\
  delta  &$8870.9$ &  $0.02$  &$0.048$  &   --  & $0.096$  &$ 0.048$
& $0.048$ \\
    phi &$11044.8$ &  $0.02$  &$0.052$  &   --  & $0.018$  &$0.0178$
& $0.052$ \\
\hline
\hline
\end{tabular}
\end{center}

* laboratory values determined by HNC95

B90: \cite{brett90}\\ 
DLP86: \cite{davis86}\\ 
J94: \cite{TiOJorg}\\ 
HNC95: \cite{TiOfel95}\\ 
L97: \cite{Langhoff97}\\ 
AP98: \cite{AP98}\\ 
\end{table}

\clearpage
\section{Figures}

\begin{figure}[ht]
\caption[]{\label{dwarfs-blue} Comparison between solar abundance
M dwarf models calculated using \amestio\ (full curves) and
\jtio\ (dotted curves) in the blue spectral region.}
\end{figure}

\begin{figure}[t]
\caption[]{\label{dwarfs-red} Comparison between solar abundance
M dwarf models calculated using \amestio\ (full curves) and
\jtio\ (dotted curves) in the red spectral region.}
\end{figure}

\begin{figure}[t]
\caption[]{\label{tio29} Comparison between $\teff = 2900\,$K;
$\logg=5.0$ and solar abundance models calculated using \amestio\
(full curves) and \jtio\ (dotted curves) based on the same set of TiO
oscillator strengths (see ``adopted'' in Table~\ref{feltab}).  The
models use otherwise identical opacities and parameters.  Each model
is fully converged, and the synthetic spectra are downgraded to a
resolution of $20\ang$.  The positions of the TiO band heads are
indicated according to table 8 of \cite{Langhoff97}.}
\end{figure}

\begin{figure}[t]
\caption[]{\label{giants-blue} Comparison between solar abundance M
giant models calculated using \amestio\ (full curves) and \jtio\
(dotted curves) in the blue spectral region.}
\end{figure}

\begin{figure}[t]
\caption[]{\label{giants-red} Comparison between solar abundance
M giant models calculated using \amestio\ (full curves) and
\jtio\ (dotted curves) in the red spectral region.}
\end{figure}

\begin{figure}[t]
\caption[]{\label{dwarf-structure} Comparison between solar abundance
M dwarf models with $\Teff=2500\K$ calculated using \amestio\ and
\jtio. The differences are calculated in the sense \amestio\ minus
\jtio\ model. The bottom panel gives the results for the Planck-mean
(full curve), J-mean (dotted curve), Flux-mean (dashed curve) and
Rosseland mean dot-dashed curve) opacities.}
\end{figure}

\begin{figure}[t]
\caption[]{\label{giant-structure} Comparison between solar abundance
M giant models with $\Teff=3000\K$ calculated using \amestio\ and
\jtio. The differences are calculated in the sense \amestio\ minus
\jtio\ model. The bottom panel gives the results for the Planck-mean
(full curve), J-mean (dotted curve), Flux-mean (dashed curve) and
Rosseland mean dot-dashed curve) opacities. }
\end{figure}

\begin{figure}[t]
\caption[]{\label{mt-compare} Comparison between solar abundance M
dwarf models calculated using \ameswater\ (full curves) and \mtwater\
(dotted curves) in the near-infrared  spectral region. Both sets of models
were calculated using AMES-TiO and iterated to convergence with their
respective parameters.}
\end{figure}

\begin{figure}[t]
\caption[]{\label{mt-structures} Comparison between solar abundance M
dwarf models with $\Teff=2500\K$ calculated using \ameswater\ and
\mtwater. The differences are calculated in the sense \ameswater\
minus \mtwater. The bottom panel gives the results for the Planck-mean
(full curve), J-mean (dotted curve), Flux-mean (dashed curve) and
Rosseland mean (dot-dashed curve) opacities. Both sets of models were
calculated using AMES-TiO and iterated of convergence with their
respective parameters.}
\end{figure}

\begin{figure}
\caption[]{\label{TiO3500} We compare the optical spectral
distribution of a NextGen model with $\Teff=3500\,$K;
$\logg=5.0$; $\mh=0.0$ (dotted line) with a model converged on the same
parameters using the AMES-TiO and H$_2$O line lists (full line).  The
positions of the TiO band heads are indicated according to table 8 of
\cite{Langhoff97}.  The region of integration of standard optical
broadbands are also shown for reference. The ``AMES atmosphere'' shows
weaker TiO bands, principally due to the smaller oscillator strengths
predicted by the Langhoff TiO model.}
\end{figure}

\begin{figure}
\caption[]{\label{H2O3500} Same as in Figure~\ref{TiO3500} for the
near-infrared portion of the spectrum. The region of integration of
the standard near-infrared broadbands are also shown for reference.
The ``AMES atmosphere'' shows stronger H$_2$O bands, especially in the
trough of the bands, i.e., at 1.6 $\mu$m and at 2.1 $\mu$m. But
little or no changes are seen in the {\bf J} bandpass region.}
\end{figure}

\begin{figure}
\caption[]{\label{VRI} The optical Cousins broad-band synthetic
photometry of solar metallicity and fixed gravity ($\logg=5.0$)
models of the NextGen grid (dotted line), AMES-MT grid (long-dash
line), and AMES grid (full line) are compared to the photometric
sample of \cite{leggett92}.  This sample contains mostly M dwarfs and
metal-depleted M subdwarfs of the solar neighborhood, and becomes scarce
in the late-type dwarf regime.}
\end{figure}

\begin{figure}
\caption[]{\label{JHK} Same as Fig.~\ref{VRI} for near-infrared
broad-band colors covering the water opacity range. Please note that
the hot star tail of the sample, near H-K=0.1, is reproduced by the
NextGen and AMES-MT models for the lower gravities predicted by
evolution models for 5 to 10 Gyrs isochrones.  Of the models shown,
only the NextGen are grain-less, which explains their curling up at the
low temperature end compared to AMES-MT models.}
\end{figure}

\begin{figure}
\caption[]{\label{IJK} Same as Fig.~\ref{VRI} and~\ref{JHK} for
broad-band colors sampling side-to-side of the SED's flux peak.}
\end{figure}

\end{document}